# Fuzzy Based Diagnostics System for Identifying Network Traffic Flow Anomalies

## Gobithasan Rudrusamy, Azrudin Ahmad, Rahmat Budiarto, Azman Samsudin, Sureswaran Ramadass

*Network Research Group, School of Computer Sciences*
*Universiti Sains Malaysia, Minden Campus*
*11800, Minden, Penang, Malaysia*
*Tel: +604-8602692, Fax: +604-6574757, E-mail: { gobithasan,azru}@nrg.cs.usm.my, { rahmat,azman,sures}@cs.usm.my*

## Abstract

*In recent years, much work has been constructed in the area of tool development in order to ease a network administrator's job. However, there lack tools to collect and process flow data efficiently. This paper discusses the usage of network traffic properties in passive network monitoring which are used in recognizing and identifying anomaly. A fuzzy based diagnostic system imbedded with properties to recognize and identify network operation anomaly intelligently along with Neural Network as tuner has been proposed in this paper.*

*This paper focuses on constructing a fuzzy system by manipulating the decoded data packets (inputs) to identify anomalies. Aspects such as the selection of a suitable fuzzy set operation and tuning it have proved to increase the reliability of the computed result. In this approach, Takagi Sugeno's Fuzzy model has been implemented. With this fuzzy model, network operation anomalies are detected in accord with the intensity of the anomaly. This model also has the capability of choosing the suitable type of alerts; log, email or sms. By incorporating the fuzzy model with neural network, network operators are able to spend more time troubleshooting faults, thus minimizing the downtime of a particular segment in a network.*

## Introduction

The idea of developing a fuzzy based diagnostics for identifying network traffic anomalies in a network operation has been a vital step to overcome problems as well as to serve as an aid in the field of network monitoring. The need for proper network monitoring tools is essential to simulate worst case or stressed network performance scenarios in order to assist in network capacity planning.

Early experiments with traffic measurement tools such as NetMon [1] and Spade [2] developed for monitoring network traffic, and MRTG [3], a popular tool used by many network operators for network traffic measurement, confirmed some similarities which had some drawbacks. Some of the known features of these tools which are also part of the reporting engine are namely to provide visual data and/or create log files which require tedious inspection of the log files. The other constraint would include indication of false alarm and/or excessive notification of an anomaly at a particular timestamp. Lastly, all the alarming and reporting facilities have a large monolithic application which is difficult to manage and configure.

The steps involved in the process of network traffic inspection are to classify and identify precise characterization of anomalous network traffic behavior hence representing it in a set of rules which make up the heart of Fuzzy System. The main benefit of this system is to reduce the volume of data that the network administrators need to analyze, thus enabling them to spend more time on tasks that require higher skill levels, such as diagnosing the cause of the problem, and fixing the fault [16].

Three classified major anomalous traffic behaviors are network operation anomalies, network abuse anomalies and flash crowd anomalies [4]. This paper focuses mainly on developing a tool to monitor data network passively in the field of network operation anomalies which include plateau behavior, network device outages and significant differences due to configuration changes. Flash crowd behavior [5] has also been looked into. The main aim of this research is to experimentally reduce some of the possible constraints of the current existing tools by introducing a fuzzy based diagnostics system in for identifying network traffic anomalies.

## Approach and Methods

### The Main Structure of the System

The main architecture of the fuzzy system is featured in Figure 1. This is a general overview of the whole system. All the fuzzy set operations depicted in the shaded boxes are defined in the following subsections of this paper.

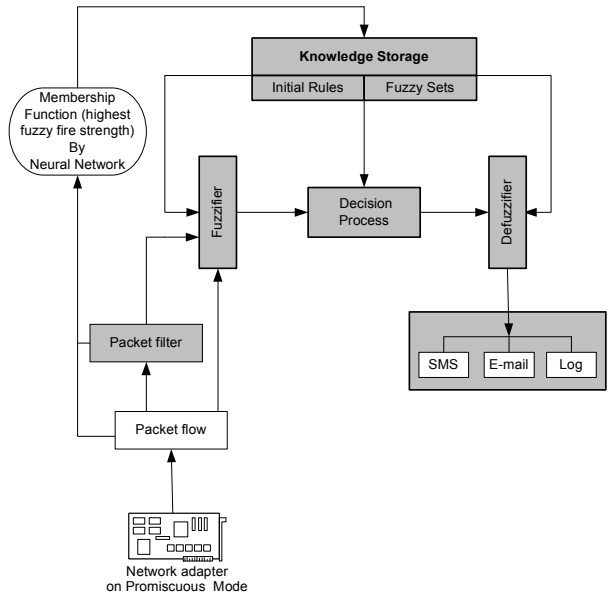

*Figure 1 – The structure of the system*

Referring to Figure 2, the system consists of two major parts which are Fuzzy Logic and Neural Network. The former is used in the decision making processes and the latter for learning processes. The implementation of both parts mentioned forms a hybrid system which has the capability of learning, adaptation and identification. However, in this paper, the focus is only given on constructing a knowledge-based Fuzzy system by manipulating the data packet in order to identify anomalies. The characteristics of the fuzzy system application in this case would be as customary, with decoded data packet as inputs.

The system consists of two modes, the 'Survey' mode and the 'Ready to Alert' mode. Each mode plays its respective function as stated in Figure 2. The functions are important since different data network conveys different pattern in packet movements in the wire. Likewise, the same data network with different segment has different characteristics. Each network has its respective peculiar network traffic curve that does not change significantly over the time [13]. In this case, Neural Network has been implemented as a tuner for fuzzy system.

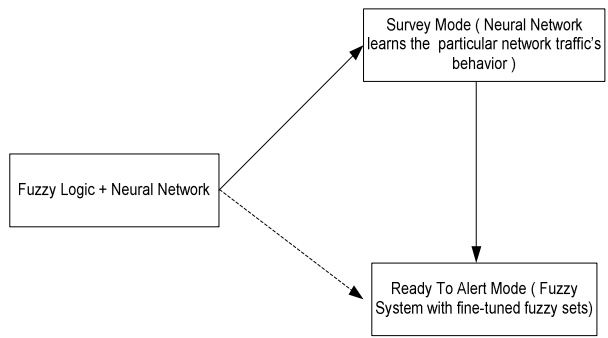

*Figure 2 – The two modes for the system.*

### Inputs from network adapter

The Packet Capture Driver is manipulated in order to have the capability to capture raw data packets from Ethernet/IEEE 802.3 technologies [6]. It sniffs all the packets in the wire, not just the packets [7] which are intended for the particular node. The received frames are decapsulated and are filtered in accord with the parameters to be set for the system.

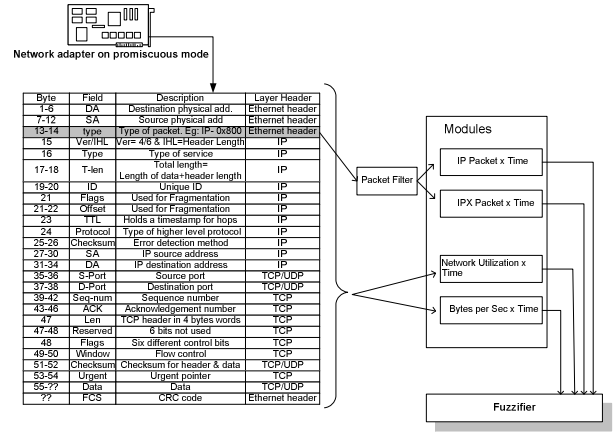

*Figure 3 – The particular fields tunneled into the fuzzifier in accord with the modules.*

### The system variables and fuzzy parameters

By capturing the essentials of the problem, the design of the process is constructed, leaving aside all the factors that could be arbitrary. In general, the simpler parameters which are kept contribute to the more understandable behavior of the system and will be more efficient in the sense of computation power consumption [8], [9], [10], [11]. In this paper, much attention is given to the aspects of selecting a suitable fuzzy set operation and tuning it, taking into consideration the reliability of the computed result.

In this system, Takagi and Sugeno's fuzzy model [10] has been implemented. This fuzzy model can be formulated as the following form:

$$R_i : \quad \text{If } \xi_1 \text{ is } A_1^i \text{ and } \ldots \text{ and } \xi_n \text{ is } A_n^i \text{ then}$$
$$y^i = a_0^i + a_1^i \cdot \xi_n + \ldots + a_n^i \cdot \xi_n \quad (1)$$

Where $R_i \, (i = 1, 2 \ldots l)$ denotes the i-th implication, $l$ is the number of fuzzy implications, whereas $y^i$ is the output from the i-th implication. Consequent parameters are $a_p^i \, (p = 0, 1 \ldots n)$ with $\xi_1, \ldots \xi_n$ as the antecedent variables and $A_p^i$ are fuzzy sets whose membership functions are denoted by the same symbols as the fuzzy values.

Figure 4 presents the standardized antecedent parameters used for the four modules stated in Figure 3. Since time is an invariant element that plays an important role in gaining pattern, it is used against type of packets, number of packets and size of packets captured respectively for a fixed timestamp.

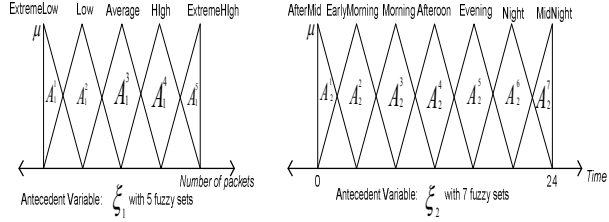

*Figure 4 – The input parameters for data network traffic diagnostics*

Out of the different fuzzy variable membership functions, the type used for this research is the triangular function. Triangular functions are used because piecewise linear functions are easy to handle with computer in the sense of storage and computations [10].

The crisp input for the first two modules represent the number of packets according to type transmitted at a particular timestamp. For the purpose of monitoring, the Internetwork Packet (IP packets) and Internetwork Packet Exchange (IPX packets) which are the majority type of packets transmitted through the Netware server in the School Of Computer Science, USM (NRG Lab) are experimented with.

NetWare is used in the lab to provide services such as transparent remote file access and distributed network services namely printer sharing [13]. Hence, it is important to watch over the IPX type of packets. The advantage about Netware is that, it is designed with IPX RIP and IPX SAP which broadcasts every 60 second for updates.

For the case of IP packets, it is used to encapsulate different type of packets such as UDP, TCP, ICMP, OSPF and IGMP. It captures a broader perspective while retaining its unique pattern for a particular data network segment.

Modules for Network Utilization and Bytes per Sec. are constructed based on the standardized antecedent parameters as in Figure 4. For rules construction, data collected from the selected network segments are investigated for patterns. The data concerning abnormal network traffic flow behaviors are analyzed thoroughly. Utilizing the available data together with knowledge expertise, the finer rules are developed producing a precise alert type with minimum overhead.

**Linguistic rules and evaluation**

The control rules are defined using the linguistic terms associated with fuzzy sets that appear in the fuzzy partitions of the domains. Figure 5 shows the initial control rule for IP Packet vs. Time module. However, there has to be different types of rules constructed for other respective modules. Extreme cases are facilitated for immediate response by the Short Message Services (SMS) and responses differ according to the fluctuation in the percentage of anomaly.

| $\xi_2$ \ $\xi_1$ | ExtremeLow | Low | Average | High | ExtremeHigh |
|---|---|---|---|---|---|
| AfterMid | SMS | IGNORE | LOG | SMS | SMS |
| EarlyMorning | SMS | IGNORE | LOG | EMAIL | SMS |
| Morning | SMS | LOG | IGNORE | IGNORE | SMS |
| Afternoon | SMS | IGNORE | IGNORE | LOG | SMS |
| Evening | SMS | LOG | IGNORE | IGNORE | SMS |
| Night | SMS | IGNORE | LOG | EMAIL | SMS |
| Midnight | SMS | IGNORE | LOG | SMS | SMS |

*Figure 5 – The initial control rules for IP Packet vs. Time module*

From Equation 1

$A_1^i \in \{ExtremeLow, Low, Average, High, ExtremeHigh\},$

$A_2^i \in \{AfterMid, EarlyMorning, Morning, Afternoon, Evening, Night, Midnight\},$

$a_0^{(A_1^5, A_2^7)}, \ldots\ldots\ldots\ldots\ldots a_2^{(A_1^5, A_2^7)} \in R$

The consequent sets are in linear form as stated in equation (1). The decision logic determines the degree to which a measured input fulfils the premise of the rule (called degree of applicability) [10]. The decision logic applies each rule $R_i$ separately. The value of equation (2) gives the degree of applicability of the premise of the rule $R_i$ for k control rules.

$$\alpha_r = \min\{\mu_{i_{1,r}}^{(1)}(x_1), \ldots, \mu_{i_{1,n}}^{(n)}(x_n)\} \quad (2)$$

It can be derived from equation (2), that rule $R_i$ implies for the measured input $(x_1, \ldots, x_n)$ the fuzzy set.

$$\mu_{x_1,\ldots,x_n}^{output(R_i)} : Y \to [0,1],$$
$$y \mapsto \min\{\mu_{i_{1,r}}^{(1)}(x_1), \ldots, \mu_{i_{n,r}}^{(n)}(x_n), \mu_{i_r}(y)\} \quad (3)$$

**The choice of the defuzzification strategy**

Given an input $\xi_1, \ldots \xi_n$ the final output of the fuzzy model, is inferred by taking the weighted average of the $y^i$'s:

$$y = \frac{\sum_{i=1}^{l} w^i y^i}{\sum_{i=1}^{l} w^i} \quad (4)$$

where $w^i > 0$, and $y^i$ is calculated for the input by consequent equation of the i-th implication, and the weight $w^i$ implies the overall truth value of premise of the i-th implication for input calculated as :

$$w^i = \prod_{p=1}^{n} A_p^i(\xi_p) \qquad (5)$$

## Result

**Comparison with existing approach**

Accurate characterizing of important classes of anomalies greatly facilitates their identification which depends on robust and timely data [13]. Some of the current best practices for identifying and diagnosing traffic anomalies consist mainly of visualizing traffic from different perspectives and from prior experience [1], [2], [3]. In general, automating the anomaly identification process has been difficult in the sense of generating precise alerts to facilitate a respective abnormal condition. There has been vast amount of research on identifying network traffic anomalies utilizing tools such as Integrated Measurement Analysis Platform for Internet Traffic (IMAPIT) which has been able to provide substantial information to detect anomalies [13]. IMAPIT includes signal analysis utility which enables network traffic data to be decomposed into its frequency components using wavelet and framelet systems. Wavelet has been used to provide means for isolating characteristics of signals via a combined time-frequency representation. This is mainly to determine isolation of short and long-lived traffic anomalies. Deviation score [13] has been used to effectively isolate anomalies and has been generalized for threshold based alerts. However, less focus has been given on the alert modes, methods of investigating and facilitating various networks.

In order to overcome this, methods of embedding intelligence in network diagnosis tool has been experimented with in this research. Fuzzy logic has been implemented to accommodate this purpose by constructing rules derived from daily and weekly traffic cycle data.

**FL based Diagnostics**

Two main groups of anomalies separated based on observed duration are long-lived and short-lived event [13]. The first group consists of flash crowd events which are long-lived event. Example of flash crowd event is the occurrence of heavy traffic due to services demand in the most extreme.

Referring to the Figure 6, the shaded area refers to flash crowd behavior of a segment in USM network due to a software release. During the flash crowd event it should be noted that both the average of packet size increment and the usage of module Byte Per Sec. enables easy detection. The height of peak of the graph in Figure 6 determines the appropriate decision to be taken based on initial rules stated in Figure 5. The four decisions in this case are ignore under normal circumstances, log when the peak deviates slightly higher than normal, email when the deviation of the peak is more and Short Message Services (SMS) when deviation is acutely high. For the case of Figure 6, the decision of periodic email would be taken.

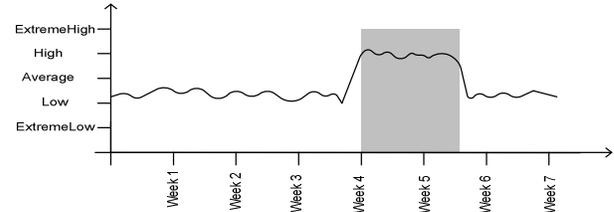

*Figure 6 – Flash Crowd Behavior based on Bytes Per. Sec. Module.*

Figure 7 shows the comparison between the normal and abnormal network traffic flow in the NRG segment. The abnormality of the flow was monitored on a different day in which power failure had occurred. This is an example of the short lived anomaly which was detected. Taking into consideration of the scenario which had a sudden and rapid fluctuation of the spike, the typical response was to alert via SMS. In the case of other factors contributing towards the abnormality of the flow, the pattern of the graph will show deviation from the standard pattern. This is clearly shown in Figure 8 whereby a network device, for instance a hub, caused a network segment to go down.

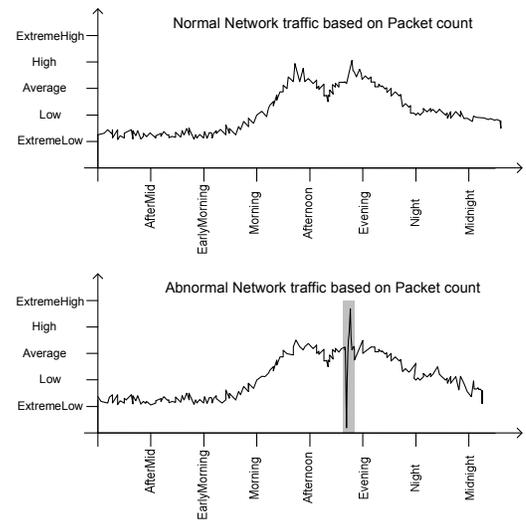

*Figure 7 – Comparison of normal and abnormal network traffic based on packet count*

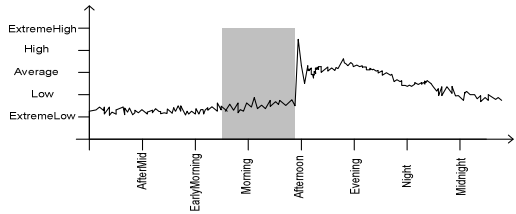

*Figure 8 – Failure of network device based on Packet count.*

Another example of a different short-lived anomaly caused by loss of the router's connectivity is shown in Figure 9. In this instance, the response of alert chosen was to SMS due to the reason that the pattern showed extremely low in packet count.

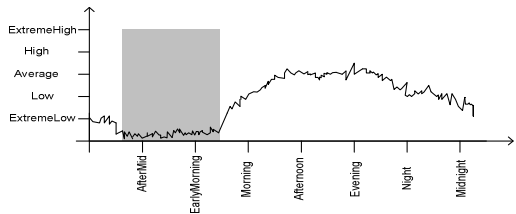

*Figure 9 – Failure of a router detected based on packet count*

The figure below shows the packet count obtained upon filtering IPX packet from Novell Netware operating system. The graphs show a comparison between the normal and abnormal pattern. The deviation of the abnormal pattern was caused by hardware failure in this case a network adapter.

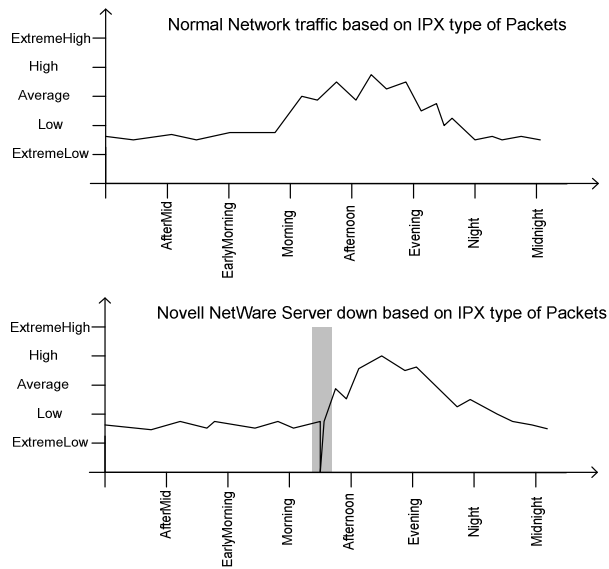

*Figure 10 – Comparison of normal and abnormal pattern for IPX type of packets.*

## Discussion

**Optimizing with Neural Network**

The fuzzy system is unable to completely function intelligently without a tuner of the parameter. Taking this into consideration, there has been means to tune the system ad hoc which is time consuming. As a step to overcome this problem, Neural Network has been introduced into the system to act as an intelligent tuner [14]. Neural network modifies the membership function when the survey mode is activated. It serves as an important function in producing the 'Δx', displacement of the vertex of fuzzy set (with the highest fuzzy strength) for the membership function in order to obtain a better control of the alerts which act as consequent parameter. By this process, it has the ability to learn a unique pattern of a particular network segment and alert accordingly.

There have been researches involving ways of optimizing neural network to define universe of discourse along with adding controlling rules for best accuracy in selecting the suitable alert type.

**Future Work**

The action of the system currently works based on a simple counter in order to avoid spamming or creating large number of alerts for a particular anomaly. In order to overcome this, the research is looking into means to stabilize the action of the alerts. A proposed method for this is a function which works as a feedback fuzzy controller system [15]. This will not only lead to controlling the type of action to be taken but is also able to control the number of the action taken. Another alternative way to tackle the problem would be to exchange the initial control rules.

The other aspect which can be explored further is to enhance the readability of anomalies by applying a variety of time-frequency analysis techniques, particularly wavelet and deviation score. These techniques consider signal variation in both high and medium frequency bands and are found to be extremely effective at isolating anomalies [13]. Utilization of techniques as such enables the achievement of two goals. One, a finer characteristics of anomalies which leads to accurate type of action and secondly, informative alerts can be generated for more efficient troubleshooting to minimize network downtime.

## Conclusion

In this research, aspects compromising of a Fuzzy based system together with Neural Network has been implemented to develop a tool which takes action intelligently. All the fuzzy operators which are fixed in order for the system to be optimized and reducing false alerts have been discussed in detail. Behavioral learning of the patterns by neural network has been explored as well to obtain more accurate alerts. The results obtained shows that number of anomalies can be detected particularly network operation anomalies. Filters at the level of network layer are implemented in order to scrutinize in accord to a particular type of packet. In this case, the IPX and IP type of packets are looked into and discussed in elaborate.

Some constraints evidently arise during this research. Due to uncontrollable network operations, some difficulty was faced when taking the statistical reading of the network traffic flow data. The other constraint arises during the process of defining a standard pattern for a particular network segment. Measures to tackle this problem were done by obtaining reading from various network segments rather than from one particular segment. Since there lacks a standardized definition for network anomalies, hence the process of characterizing the anomaly is carried out in a purely empirical approach. The proper advancement of tool for pattern gathering will give a breakthrough in the area of anomaly identification.

Finally the exploration of ideas in this paper augments the existing threshold based alerts. It is hoped that a system incorporating intelligence in software based tools is developed to help administrators [17]. Since data network has become a vital aspect in the era of information technology, the downtime of a network should be minimized. In this sense, these tools are helpful enough to passively notify network administrator regarding the health of a network from time to time.

## Acknowledgment

The authors wish to thank University Sains Malaysia for providing the research grant which was utilized in this research.

## References

[1] Ramadass, S. 2001. Network Monitor. In Proceedings of Asia Pacific Advanced Network Conference, 2001, 40-44. Penang, Malaysia.

[2] Hoagland, J. and Staniford, S. Statistical Packet Anomaly Detection Engine URL http://www.silicondefense.com/software/spice/index.htm

[3] Oetiker, T. and Rand, D. 16 October 2002. Multi router Traffic Grapher URL http://www.mrtg.cz/

[4] Barford, P. and Plonka, D. June 2001. Characteristics of Network Flow Anomalies, In Proceedings of ACM Internet measurement Workshop ACM SIGCOMM 2001. San Francisco.

[5] Jung, J., Krishnamurthy, B. and Rabinovich, M. 2002. Flash Crowds and Denial of Service Attacks: Characterization and Implications for CDNs and Web Sites. URL http://citeseer.nj.nec.com/cache/papers/cs/25742/http:zSzzSzwww.research.att.comzSz~balazSzpaperszSzwww02-fc.pdf/jung02flash.pdf

[6] Degioanni, L., Risso, F., Varenni, G. and Viano, P. 8 August 2002. WinPcap: The Free Packet Capture Architecture for Windows. URL http://winpcap.polito.it/

[7] Forouzan, B.A. eds. 2000. *TCP/IP Protocol Suite*. McGraw-Hill.

[8] Cox, E. eds. 1998. *The Fuzzy Systems Handbook Second Edition*. Chappaqua, New York. Academic Press.

[9] Yan, J., Ryan, M. and Power, J., A. eds. 1986. *Using Fuzzy Logic*. Reading, Mass. Addison-Wesley.

[10] Kruse, R, Bebhardt, J. and Klawonn, S. eds. 1993. *The foundation of Fuzzy Logic*. John Wiley & Sons.

[11] Negoita, CV. eds. 1985. *Expert Systems and Fuzzy System*. The Benjamin/Cummings Publishing Company Inc.

[12] Ramos, E., Schroeder, AL. and Simpson, L. eds. 1992. *Data Communication and networking fundamentals using Novell NetWare*. Reading, Mass. Addison-Wesley.

[13] Barford, P., Kline, J., Plonka, D. and Ron, A. November 2002. A Signal Analysis of Network Traffic Anomalies. In Proceedings of ACM SIGCOMM Internet Measurement Workshop 2002. Marseiilles, France.

[14] Herrmann, C.S. August 1995. A Hybrid Fuzzy-Neural Expert System for Diagnosis. In Proceedings of the International Joint Conference on Artificial Intelligence, Montreal, Canada.

[15] Singh, S. and Steinl, M. 24-25 October 1996. Fuzzy Search Techniques in Knowledge-Based System. In Proceedings of the Sixth International Conference on Data and Knowledge Systems for Manufacturing and Engineering (DKSME '96). Tempe, Arizona.

[16] Leckie, C. 1995. Experience and Trends in AI for Network Monitoring and Diagnosis. In Proceedings of the International Joint Conference on Artificial Intelligence Workshop on AI in Distributed Information Networks. Montreal, Canada.

[17] Meystel, A. and Messina, E. 17-19 July 2000. The Challenge of Intelligent Systems. In Proceedings of the 15th IEEE International Symposium on Intelligent Control (ISIC 2000). Rio Patras, Greece.